\documentclass[a4paper,12pt]{article}
\usepackage[a4paper,width=150mm,top=25mm,bottom=25mm,bindingoffset=6mm]{geometry}

\usepackage[english]{babel}

\usepackage{latexsym}
\usepackage{graphics}
\usepackage{graphicx} 
\usepackage{epsfig}
\usepackage{bm} 
\usepackage{amsmath}
\usepackage{amssymb}
\usepackage{amsthm}
\usepackage{amsfonts}
\usepackage{dcolumn}
\usepackage{float}
\usepackage{color}
\usepackage{epstopdf}
\usepackage[svgnames]{xcolor}
\usepackage{braket}
\usepackage{multirow}
\usepackage{rotating}
\usepackage{algorithmic}
\usepackage{textcomp}
\usepackage{cite}
\usepackage{url}
\usepackage{pifont} 
\usepackage[bookmarks=false,colorlinks]{hyperref}
\hypersetup{hidelinks,colorlinks=true,allcolors=DarkBlue}
\usepackage{appendix}

\usepackage{ulem} 
\usepackage{lineno}

\definecolor{pink}{RGB}{219, 48, 122}

\def\mod\text{mod~}

\def\yes{\ding{51}}
\def\no{\ding{55}}

\theoremstyle{remark}

\begin{document}


\title{\bf\Large Asymmetric adaptive LDPC-based information reconciliation for industrial quantum key distribution}
\author{\sc Nikolay Borisov, Ivan Petrov and Andrey Tayduganov}
\date{} 
\maketitle

{\centering \sl\small \noindent
National University of Science and Technology MISiS \\ Moscow 119049, Russia
\par}


\vskip4cm
\begin{abstract}
    We develop a new approach for asymmetric LDPC-based information reconciliation in order to adapt to the current channel state and achieve better performance and scalability in practical resource-constrained QKD systems. The new scheme combines the advantages of LDPC codes, \textit{a priori} error rate estimation, rate-adaptive and blind information reconciliation techniques. We compare the performance of several asymmetric and symmetric error correction schemes using real industrial QKD setup. The proposed asymmetric algorithm achieves significantly higher throughput, providing a secret key rate very close to the symmetric one in a wide range of error rates. Thus, our approach turns out to be particularly efficient for applications with high key rates, limited classical channel capacity and asymmetric computational resource allocation.
    \begin{description}
        \item[Keywords:] quantum communication, quantum key distribution, QKD, information reconciliation, adaptive error correction, LDPC.
    \end{description}
\end{abstract}

\pagebreak

\section{Introduction}

Quantum key distribution (QKD) systems are considered as unconditionally secure trusted couriers for symmetric-key encryption of classical communication. 
According to the common principle of QKD, the so-called ``quantum'' part of a protocol is followed by the classical post-processing procedure in order to distill the raw key copies and form a common secret key for both participants.
This post-processing procedure consists of the following basic steps: sifting, error correction, parameter estimation, privacy amplification and session authentication \cite{Kiktenko16}. In particular, the error correction (EC) process is executed in order to find and correct all incompatible bits using a classical public channel but still keeping most of the key data unrevealed. This step is thought to be the most computationally complex and time-consuming of the entire procedure \cite{Mehic20}, and therefore includes a potential for optimization. It is also the most secret-key-reducing part \cite{Brassard1994, Elkouss10secure}.

The EC procedure, often referred to as information reconciliation (IR), can be implemented in QKD using various EC techniques \cite{Brassard1994, Martinez15demyst_cascade,Vatta11,Elkouss09,Kiktenko20polar,Niemiec19}. Among them, the low-density parity-check (LDPC) codes \cite{Gallager} are well studied and widely applied in modern telecommunication systems. 
The main advantage of the LDPC codes is the possibility to reach information rates arbitrarily close to the Shannon limit for a wide variety of channels \cite{Mackey99}. That possibility can be implemented by satisfying several basic conditions, such as efficient decoder design, parity check matrix construction, and apropos code rate adaptation. 

The straightforward LDPC-based EC is asymmetric since the syndrome decoding process is much more computationally complex than the encoding one. In a scheme where the transmitter (Alice) performs encoding and the receiver (Bob) is decoding, Alice's computer is idle most of the time and the entire throughput is determined by the decoding speed of Bob's computer. Therefore, to avoid this asymmetric and inefficient use of computational resources, some modern industrial point-to-point QKD solutions implement a bi-directional \cite{Dixon2014high,Yuan18toshiba} or symmetric blind IR approach~\cite{Kiktenko17sym}.
The first approach is based on the parallel processing of different sifted key frames by Alice and Bob, achieving a Mbps key throughput. The second one combines the advantages of LDPC and the interactivity of the Cascade IR protocol in the correction of the same frame, achieving fail-resistant performance and high information efficiency simultaneously. 
These highly-efficient reconciliation schemes, however, remain practical only if both Alice and Bob possess sufficiently powerful computing devices.

Nowadays, the latest commercially available QKD devices (manufactured by e.g. ID Quantique, QuantumCTek, Toshiba) are designed as 19-inch rack modules, having the same size specifications for both transmitter and receiver. However, in some practical QKD schemes and applications one would like to minimize the size and cost of the transmitter device and as a consequence reduce its computational load. For instance, in an urban star-topology quantum network with one powerful receiver server connecting multiple users, it would be ideal in the future to develop a user's transmitter device of the PCI-e card form factor to be installed inside a normal personal computer. In order to save the processor's computation resources and power consumption such a transmitter has to execute minimum complex EC operations. Another example is the satellite QKD in which the satellite is a transmitter playing a role of a trusted node \cite{Liao17satellite}. Apparently, for the same reasons such a device requires minimization of the load on its computing module and on the service channel, preserving efficient performance under unstable quantum channel conditions. One can also consider hand-held devices \cite{Vest22hand_held}. Therefore, for such applications, one has to reconsider the EC workflow taking into account the inequality of computational resources of transmitter and receiver. The design must be aimed to provide a real-time post-processing service for such asymmetric QKD systems.

In this work, we consider the decoy-state BB84 protocol \cite{Hwang03,Lo05,Ma05,Trushechkin17} and revisit the asymmetric LDPC-based reconciliation \cite{Elkouss11,Martinez12blind,Dixon2014high,Mao2019high,Mao2021high,Gao19multi_matrix}, in which the syndrome decoding is performed only on Bob's side, by modifying and improving the existing symmetric blind IR scheme~\cite{Kiktenko17sym}. 
We develop a novel rate-adaptive algorithm that employs a new optimal code rate selection approach based on our \textit{a priori} quantum bit error rate estimation method. It is shown that the knowledge of precise error rate value and the proper code rate choice is crucial for high-performance EC. We also propose a different additional round organization rule that allows a direct code efficiency control without round number limitation.
Using the simulated and real experimental data we demonstrate that the new asymmetric scheme is able to achieve almost the same efficiency as symmetric one, keeping the low failure probability and time consumption. The performance of our scheme on data is also compared with two blind asymmetric schemes, proposed in \cite{Martinez12blind,Liu20}, that use different bit disclosure rules during additional rounds.


The paper is organized as follows. In Section~\ref{sec:basics_IR} we review the basics of the IR procedure, particularly focusing on asymmetry-related solutions. In Section~\ref{sec:scheme} we discuss the details of the code rate scheme adaptation changes that have to be made in order to save a satisfactory overall efficiency parameter of the cryptosystem. In Section~\ref{disc:sim} we compare the asymmetric and symmetric approaches using some set of benchmarks. We summarize our results in Section~\ref{sec:conclusion}.

\section{Information reconciliation with LDPC codes}\label{sec:basics_IR}

In the BB84 QKD protocol \cite{BB84} the sifted keys of Alice and Bob, made out of raw keys by rejecting events with incompatible bases, are not 100\% identical and contain quantum bit errors that must be found and corrected. After correction and subsequent verification, the key passes to the privacy amplification step -- a special contraction of the verified key with 2-universal hash functions into a shorter unconditionally secure secret key in order to minimize information of potential eavesdropper (Eve) to an arbitrarily low value.
For practical reasons, the sifted key is accumulated in blocks of fixed size $\ell_\text{block}$. In order to minimize the effect of statistical fluctuations on the final secret key length evaluation and consequently obtain a less conservative result, $\ell_\text{block}$ has to be of the order of $10^6$ and larger. Since such bit string size is too large for high-speed and efficient LDPC-based EC, the block is split into a number of smaller subblocks of length $\ell_\text{subblock}\sim10^3-10^5$ bits and EC is performed on each subblock separately.

Each subblock correction starts with an \textit{a priori} quantum bit error rate (QBER) estimation. The straightforward approach is to disclose and compare publicly a random sample of the sifted key. Apparently, the disclosed bits have to be discarded by Alice and Bob.
To avoid such excessive key consumption, the estimation can be done by analyzing only indirect information about errors, such as polarization drift~\cite{Muga11qber_polarization} for polarization-encoding protocol or LDPC syndrome~\cite{Treeviriyanupab14qber,Kiktenko18qber_est} for general discrete variable QKD protocol.

Originally, the LDPC codes were designed for a one-way EC scheme \cite{Mackey99}. In this scheme Alice selects an appropriate code of rate $R$ that fits the quantum channel capacity and determines the corresponding sparse parity-check matrix $H_R$, calculates the syndrome $\bm{s}_A$ of length $(1-R)\ell_\text{subblock}$ from her message (key) $\bm{x}_A$,
\begin{linenomath}
\begin{equation}
    \bm{s}_A = H_R \bm{x}_A ~(\text{mod}\,2) \,,
\end{equation}
\end{linenomath}
and sends it to Bob via a classic authenticated channel. Bob in turn computes the syndrome $\bm{s}_B=H_R\bm{x}_B$ from his sifted bit string $\bm{x}_B$ and performs Alice's syndrome decoding that can be interpreted as searching for error pattern $\bm{e}$ such that
\begin{linenomath}
\begin{equation}\label{eq:H_Re}
    H_R\,\bm{e} ~(\text{mod}\,2) = \bm{s}_A\oplus\bm{s}_B \,.
\end{equation}
\end{linenomath}
Solving \eqref{eq:H_Re} Bob corrects his bit string: $\bm{x}_{B}\to\bm{x}_{B}\oplus\bm{e}\equiv\bm{x}_A$. If the solution is not found within a limited period of time, the decoding is failed.

The efficiency of LDPC algorithms can be further increased using the rate-adaptive~\cite{Elkouss10secure,Elkouss11} and blind reconciliation \cite{Martinez12blind} methods. To adapt the code rate $R$ more precisely to the current quantum channel state, we use the puncturing (and shortening) technique \cite{Ha,Yazdani}. It is based on the idea of extending subblock (payload data) to a new EC unit, here and thereafter called \textit{frame}, by inserting additional noise (punctured bits) into $\bm{x}_{A,B}$.
Then the blind IR can be applied to increase the EC success probability. In this approach, the accurate \textit{a priori} QBER estimation and initial choice of $R$ are considered to be unnecessary. Instead, if Bob reports his basic round decoding failure, during additional rounds Alice discloses some fraction of her punctured bit values and Bob makes another decoding attempt, this time having more information about the frame. The resulting efficiency deteriorates with the number of disclosed bits and the number of additional rounds. Consequently, the frame correction running time and the load on the classical channel increase, but the EC success probability increases.

The number of additional rounds is limited by the amount of information to be revealed and the revealing strategy. In the original blind reconciliation scheme \cite{Martinez12blind}, during every additional round Alice discloses the fixed number of punctured bits, $d_k=p/N_\text{add}^{\max}$, where $p=\alpha\ell_\text{frame}$ is the total number of punctured bits in the frame, defined by empiric code rate adaptation parameter $\alpha$, and $N_\text{add}^{\max}$ is the maximum allowed number of additional rounds. Another way, proposed in an attempt to make the decoder converge faster, is the linearly increasing with iteration number $k$ step $d_k=k\delta$ ($k\geq1$) where $\delta$ is determined empirically \cite{Liu20}.

In symmetrically organized reconciliation \cite{Kiktenko17sym} Alice and Bob share their syndromes and both perform belief-propagation decoding of \eqref{eq:H_Re}. In case of failure, they compute the log-likelihood ratio (LLR) for every bit in the frame and then disclose to each other only the bits with minimal LLR. To increase the decoding success probability, both punctured and payload bits are allowed to be disclosed (of course, such payload bits are excluded from the final secret key).
Then Alice and Bob refresh their EC frames using new data and perform another decoding attempt. In \cite{Kiktenko17sym} the following heuristic rule is used to determine the disclosed data amount: $d_k(R)=\lceil \ell_\text{frame}(0.028-0.02 R) \beta\rceil$, $\beta\in\{0.5,1\}$. In case of unsuccessful decoding, the process is stopped if all bits are disclosed, or the frame correction time budget is over. The authors of \cite{Kiktenko17sym} compare the blind approach with the rate-adaptive regime under the assumption that the QBER level is known and perform simulation that shows better efficiency and number of iterations.

The symmetric strategy shows itself to be highly efficient, but for the asymmetric scheme Alice has no intermediate decoder parameters such as positions of bits with minimal LLR, so she cannot disclose this auxiliary data to Bob blindly anymore. Using rather different ideas couple of solutions were proposed in Refs.~\cite{Gao19multi_matrix,Mao2021high}. In this work we step aside from the blind reconciliation principle and develop further the additional rounds' organization strategy that preserves the efficiency of symmetric IR in presence of limited computational resources.

\subsection*{IR performance metrics}

Since Eve can extract some partial useful information from syndromes and other data, exchanged via the classic channel during the IR communication rounds, this potential leakage has to be subtracted and taken into account in the final key length estimation. In order to quantify the EC efficiency and estimate the amount of disclosed information about the key, the following metric is introduced~\cite{Elkouss11}:
\begin{linenomath}
\begin{equation}
    f_\text{ec} =  \frac{\ell_\text{syndrome} - p + d}{(\ell_\text{frame} - p - s) h_2(E_\mu)} \,,
    \label{eq:f_ec}
\end{equation}
\end{linenomath}
with $\ell_\text{syndrome}=(1-R)\ell_\text{frame}$ and $p+s=\ell_\text{frame}-\ell_\text{subblock}=\alpha\ell_\text{frame}$ where
\begin{itemize}
    \item $E_\mu$ -- average signal pulse QBER of a subblock
    \item $p$ -- number of punctured bits in a frame
    \item $s$ -- number of shortened bits in a frame
    \item $d=\sum_k d_k$ -- total number of disclosed bits in additional rounds 
    \item $h_2(x)=-x\log_2x-(1-x)\log_2(1-x)$ -- Shannon binary entropy.
\end{itemize}
This metric represents the ratio of the information content, used to reconcile one data frame, over the minimal information theoretically required~\cite{Elkouss10secure}. Thus, large $f_\text{ec}$ implies less efficient EC. Large $f_\text{ec}$ also indicates the process interactivity. Note that $f_\text{ec}$ can not approach values less than 1 due to Shannon's limit. In this way, for purely theoretical studies the average information leakage per successfully corrected subblock can be estimated as $\ell_\text{subblock} f_\text{ec} h_2(E_\mu)$.

Another important code quality metric that also affects the secret key generation rate is the frame error rate (FER) -- the frame decoding failure probability. Taking into account FER, the modified formula for the average secret key length from \cite{Trushechkin17} can be written as follows,
\begin{linenomath}
\begin{equation}
    \ell_\text{sec} \simeq \ell_\text{block} (1 - \text{FER}) \big\{ \kappa_1^l [1 - h_2(E_1^u)] -f_\text{ec} h_2(E_\mu) \big\} \,,
    \label{eq:l_sec}
\end{equation}
\end{linenomath}
where $\kappa_1^l$ is a lower bound on the fraction of bits in the verified key obtained from single-photon pulses, and $E_1^u$ is an upper bound on the fraction of errors in such positions in the sifted keys (for their estimations see e.g. \cite{Trushechkin17,Zhang17}). The trade-off between low(high) $f_\text{ec}$ and high(low) FER is the main objective of this IR research.

And finally the last important factor is the CPU load. As already mentioned above, the LDPC decoder complexity basically depends on the frame length $\ell_\text{frame}$. Hence, the processing time consumption is linearly dependent on the total number of decoding iterations that can be used to analyze the load of Bob's module. Therefore, for practical QKD applications even more crucial performance criterion is the average secret key generation rate which can be estimated as
\begin{linenomath}
\begin{equation}
    R_\text{sec} = \frac{\ell_\text{sec}}{\tau} \,,
    \label{eq:R_sec}
\end{equation}
\end{linenomath}
where $\tau$ is the time needed to produce a secret key of length $\ell_\text{sec}$ or equivalently the overall block generation and post-processing time. In this way, Eqs.~\eqref{eq:l_sec} and \eqref{eq:R_sec} can be used as the main benchmarks when comparing various EC schemes with different $f_\text{ec}$, FER and number of additional EC rounds.

\section{Adaptive code rate method for asymmetric blind reconciliation}\label{sec:scheme}

\begin{figure}[p!]\centering
    \includegraphics[width=0.8\textwidth]{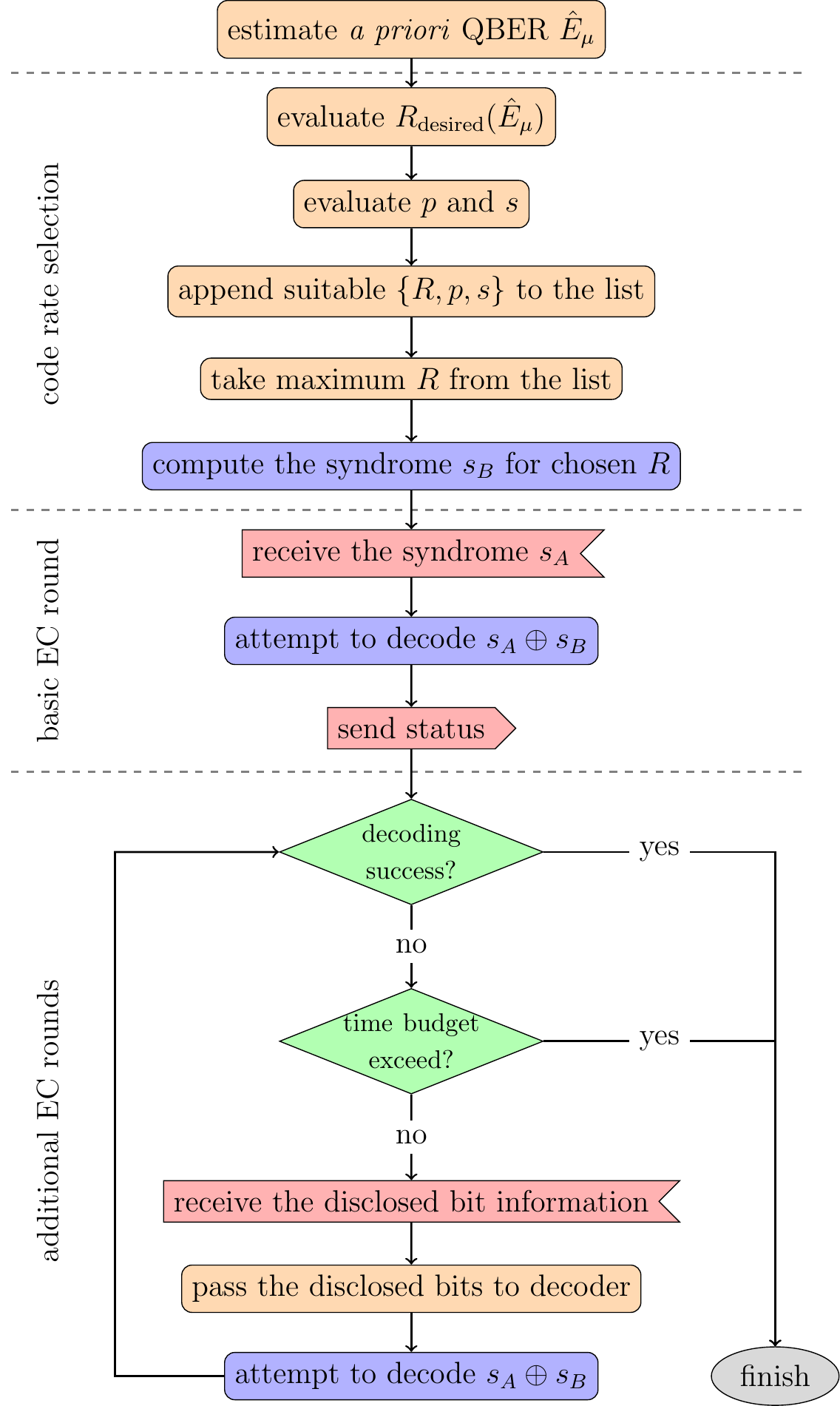}
    \caption{\footnotesize Activity diagram of the asymmetric information reconciliation process on Bob's side.}
    \label{fig:activity_diagram}
\end{figure}

In this section we describe the proposed asymmetric algorithm, schematically shown in Figure~\ref{fig:activity_diagram}, that contains three key steps explained below. Before going into details, let us first list the used basic IR parameters and tools.

In our study the appropriate frame length is chosen to be $\ell_\text{frame}=32\,000$ bits. Then the key subblock length is computed as 
\begin{linenomath}
\begin{equation}
    \ell_\text{subblock} = \ell_\text{frame} (1 - \alpha) \,.
\end{equation}
\end{linenomath}
Using $\alpha=0.15$ \cite{Kiktenko20polar} one obtains $\ell_\text{subblock}=27\,200$ bits. Since the post-processing block size has to be at least of the order of $10^6$, we take $\ell_\text{block}=50\ell_\text{subblock}=1.36\times10^6$.

In order to reduce the impact of error bursts on the decoding process and to randomize the locations of errors, we apply the interleaving technique \cite{Mehic20}. Alice and Bob simultaneously reorder bits in the subblock according to the permutation law, determined by two synchronized pseudo-random number generators based on Mersenne Twister.

The LDPC matrices $\{H_R\}$ are generated for the code pool $R\in\{0.5,0.55,\dots,0.9\}$ with the Progressive Edge-Growth algorithm~\cite{hu01} and Tanner graph node degree distributions described in \cite{Elkouss09}. The values of shortened and punctured bits are defined by pseudo-random and true number generators respectively (see \cite{Kiktenko17sym} for detailed information). The untainted puncturing technique of proper punctured bit position choice is also used \cite{Elkouss12untained}.

The Sum-Product decoder \cite{Mackey99} is the popular belief propagation LDPC syndrome decoder. However, it employs rather heavy computational operations and thus is not efficient enough for high-speed data processing. Therefore, we apply its effective approximation -- the variable-scaled Min-Sum decoder \cite{Emran14} with the scaling parameter step equal to $12.5$, which gives the best efficiency in our tests.

\subsection{\textit{A priori} QBER estimation}\label{sec:scheme_QBER}

The EC of a new frame starts with the LDPC code rate choice based on the \textit{a priori} error rate estimation. Although the blind rate-adaptive reconciliation is supposed to work without exact knowledge of QBER \cite{Martinez12blind}, it remains rather sensitive to the initial code rate value. The main idea of the blind reconciliation is to use an LDPC code of fixed rate that can be adapted by iterative disclosure of punctured bits.
Therefore, in order to choose the optimal code rate, in this work we propose to estimate the current \textit{a priori} QBER using the \textit{a posteriori} QBER information from the previously corrected and verified frames. We also take into account the non-zero frame error rate (FER), caused by either LDPC code imperfections or unexpected QBER fluctuations, and propose a simple feedback loop. 

In our scheme the \textit{a priori} QBER for arbitrary $i$-th frame $\hat{E}_\mu^{(i)}$ is estimated by the exponential moving average of the previous verified frame, $\text{EMA}^{(i-1)}$, defined iteratively as
\begin{linenomath}
\begin{equation}
    \text{EMA}^{(j)} =
    \begin{cases}
        E_\mu^{(i-6)} \,, \quad j=i-6 \,, \\
        \gamma E_\mu^{(j)} + (1-\gamma) \text{EMA}^{(j-1)} \,, \quad i-5 \leq j \leq i-1 \,,
    \end{cases}
\end{equation}
\end{linenomath}
with the empirical smoothing factor $\gamma=0.33$. The exponential weights lead to a more optimal code rate choice from the pool in case of gradual QBER variation, while the average value smoothes possible sporadic error bursts and therefore results in stable EC performance.

Nevertheless, the EMA method does not allow to detect and quickly adapt $R$ to a sudden significant leap of QBER level. Therefore, we check the presence of error bursts by analysing the set of weak decoy pulse QBERs of the block, $\{E_{\nu_1}^{(1)},\dots E_{\nu_1}^{(50)}\}$. Since decoy pulses are not used for the key formation, Alice can safely send a string of decoy bit values to Bob who compares it with his own one and computes decoy QBERs straightforwardly.
We set the following condition: if $|E^{(i)}_{\nu_1}-\mathbb{E}[E_{\nu_1}]|\geq3\sigma[E_{\nu_1}]$, where $\mathbb{E}[E_{\nu_1}]$ and $\sigma[E_{\nu_1}]$ are the mean value and the standard deviation respectively, the sporadic error burst is detected. Using the simplest theoretical model for QBER prediction e.g. from Ref.~\cite{Ma05}, one can show that $E_\mu$ and $E_{\nu_1}$ have very similar behaviour and that $\mathbb{E}[E_\mu]\leq \mathbb{E}[E_{\nu_1}]$ due to the $\mu>\nu$ condition. Thus, we can use $\hat{E}_\mu^{(i)}=E_{\nu_1}^{(i)}$ as an upper bound for the signal \textit{a priori} QBER instead of the EMA estimation. 

\begin{figure*}[t!]\centering
    \includegraphics[width=0.99\textwidth]{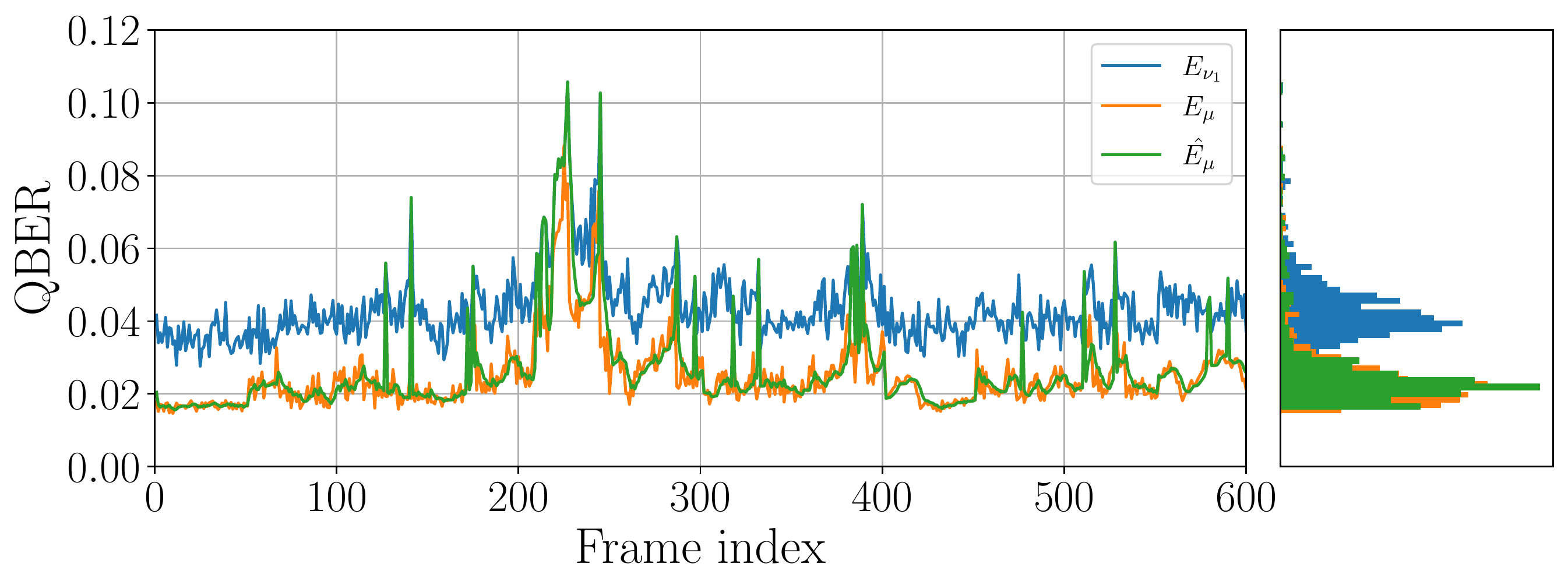}
    \caption{Example of experimental weak decoy ($E_{\nu_2}$), real ($E_\mu$) and estimated \textit{a priori} ($\hat{E}_\mu$) signal QBER. The data was generated with QKD devices by QRate for the 20\,dB quantum channel.}
    \label{fig:qber-in-time}
\end{figure*}

After the $i$-th frame correction the verification step is performed, where we propose a simple performance control rule: in case of frame verification failure the EMA is calculated with penalty value $\hat{E}_\mu^{(i)}=0.5$. Such a feedback loop provides a temporal decrease of the algorithm efficiency, increasing the probability of successful EC of the next frame. 
The workflow of our \textit{a priori} QBER estimation is shown in Figure~\ref{fig:qber-in-time}. One can observe high consistency of $E_\mu$ and its estimation $\hat{E}_\mu$ even in presence of the error burst from $2\%$ up to $8\%$ values of $E_\mu$ (frames 200-250).

\unskip

\subsection{Code rate selection}\label{sec:scheme_rate}

It is crucial for the entire algorithm to set up a proper initial efficiency value of $f_\text{ec}$ before the code rate is chosen. We use $f_\text{start} = 1.15$ as an empirical optimum. In this way we can directly control the reconciliation scheme efficiency. 
Taking into account the LDPC code's imperfection, the desired code rate is defined by Shannon's capacity of the binary symmetric channel,
\begin{linenomath}
\begin{equation}
    R_\text{desired} = 1 - f_\text{start} h_2(\hat{E}_\mu) \,.
\end{equation}
\end{linenomath}

If $0.5\leq R_\text{desired}\leq0.9$, then for every code rate $R$ from the pool $\{0.5,0.55,\dots,0.9\}$ the total numbers of punctured ($p$) and shortened ($s$) bits are estimated as 
\begin{linenomath}
\begin{equation}
\begin{split}
    p &= \ell_\text{frame} \lceil1 - R - (1 - \alpha) f_\text{start} h_2(\hat{E}_\mu)\rceil \,, \\
    s &= \alpha \ell_\text{frame} - \, p \,.
\end{split}
\end{equation}
\end{linenomath}
This list of sets $\{R, p, s\}$ is sifted considering the following conditions: 
\begin{linenomath}
\begin{equation}
    p,s \geq 0 \,, \quad p \leq p_R \,, \quad \hat{E}_\mu < t_R \,.    
\end{equation}
\end{linenomath}
Here $t_R$ is the error rate threshold defined in \cite{Elkouss09}, $p_R$ is the maximum amount of punctured bits calculated by untainted puncturing technique \cite{Elkouss12untained}. The rest of appropriate sets $\{R, p, s\}$ forms a list, from which the algorithm chooses the one with maximum $R$.

For very high/low QBER the list turns out to be empty. In this case the algorithm chooses either $\{0.5,0,\alpha \ell_\text{frame}\}$ if $R_\text{desired}\leq R_\text{min}=0.5$, or $\{0.9,p_{R_\text{max}},\alpha \ell_\text{frame}-p_{R_\text{max}}\}$ if $R_\text{desired}\geq R_\text{max}=0.9$.

\subsection{Additional correction rounds} \label{sec:scheme_rounds}

Next, we modify the scheme of additional rounds organization. If the basic decoding round does not converge successfully, Bob reports to Alice about the occurred fail, and Alice in return starts disclosing punctured node values. Since the punctured nodes are generated true-randomly, their values' disclosure eliminates rather a high amount of uncertainty for Bob's decoder, and hence with high probability these nodes have the smallest LLR values. If all punctured bits are already disclosed but the decoding is still unsuccessful, Alice continues additional rounds by disclosing pseudo-randomly chosen payload bits. In general, provided that the previous $k-1$ rounds failed, in the next $k$-th round the number of disclosed punctured/payload bits is calculated according to our rule,
\begin{linenomath}
\begin{equation}
    d_k = \bigg| \displaystyle\frac{\ell_\text{syndrome} - p + \sum_{l=0}^{k-1}d_l}{(\ell_\text{frame} - p - s) h_2(\hat{E}_\mu)} - f_k \bigg| \ell_\text{frame} \hat{E}_\mu \,, \quad k\geq1 \,,
    \label{eq:d_k_rule}
\end{equation}
\end{linenomath}
with $d_0=0$ and $f_k=f_\text{start}+0.03k$. The additional rounds take place until the successful decoding result, or in case of continuous fails, either the frame correction time budget or the maximum allowed number of iterations ($N_\text{add}^{\max}$) exceeds its limit. In the latter case Bob reports his failure status to Alice and both sides discard the corresponding subblock from the block.
We evaluate the time budget out of timeouts for data transfer operations, i.e. based on the Quality of Service (QoS) of the classic channel (main factor), sifted key generation rate and hardware computational resources, which results in a value of the order of milliseconds.

\section{Simulation and experimental results} \label{disc:sim}

\begin{figure*}[t!]\centering
    \includegraphics[width=0.45\textwidth]{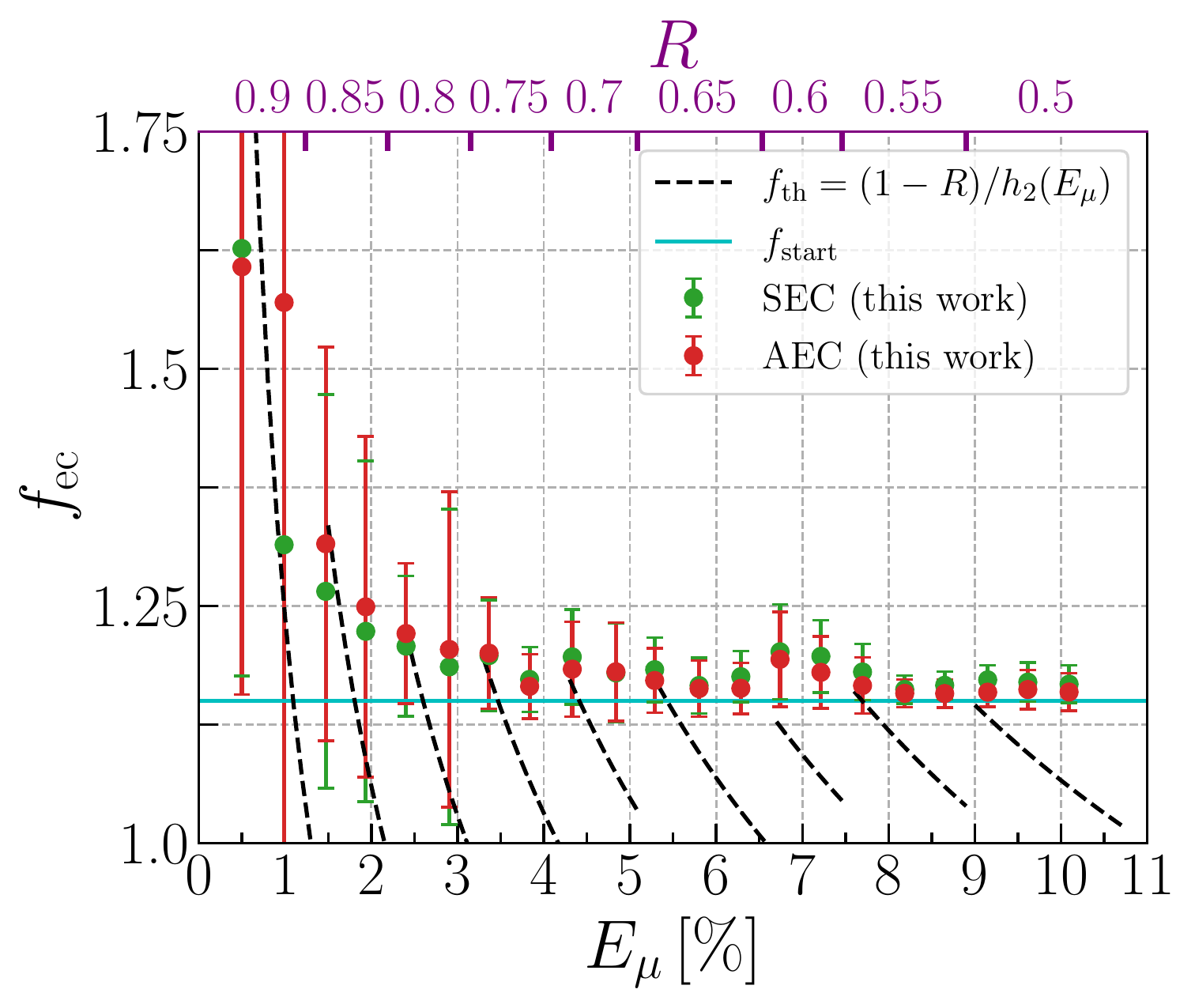}
    \includegraphics[width=0.45\textwidth]{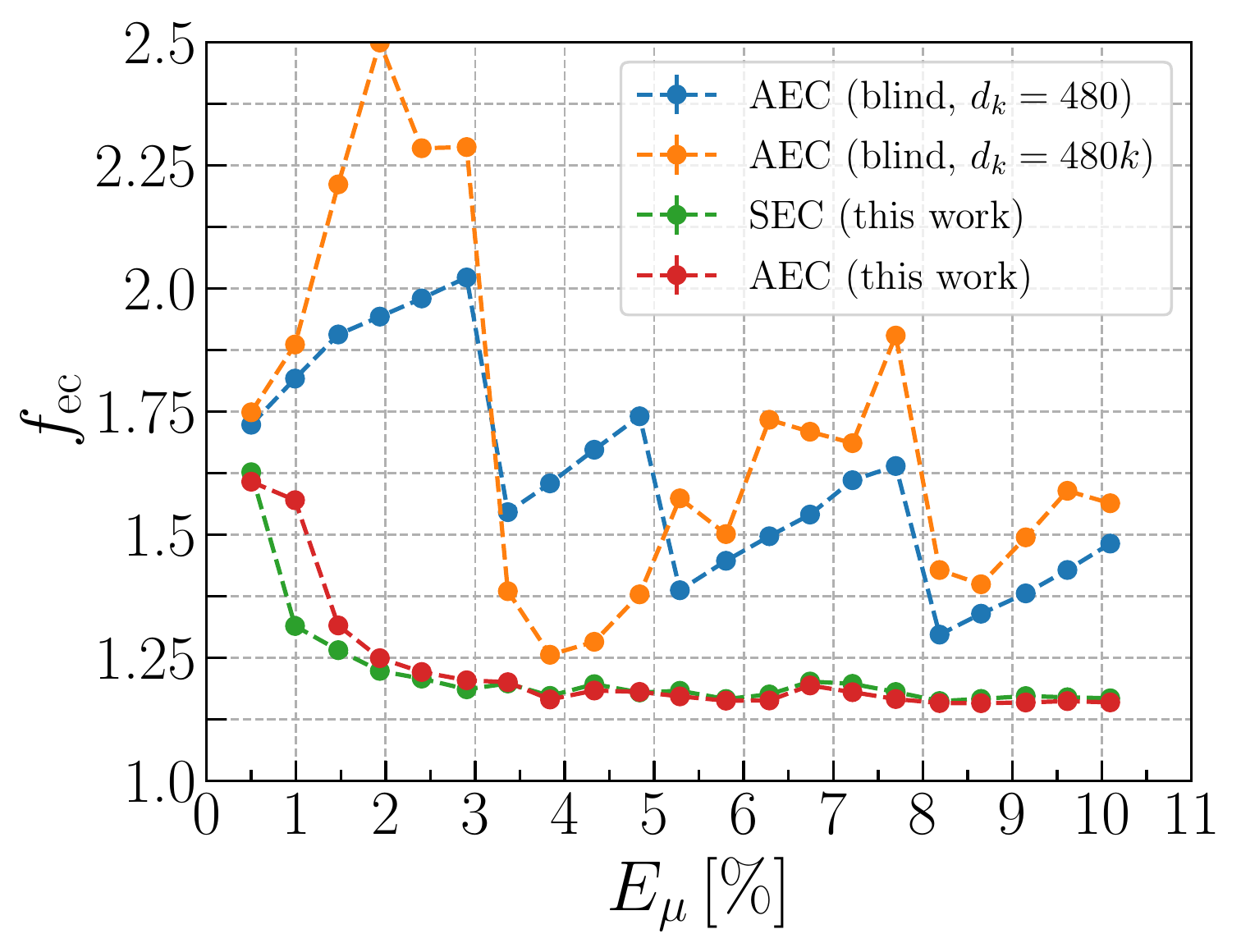}
    \includegraphics[width=0.45\textwidth]{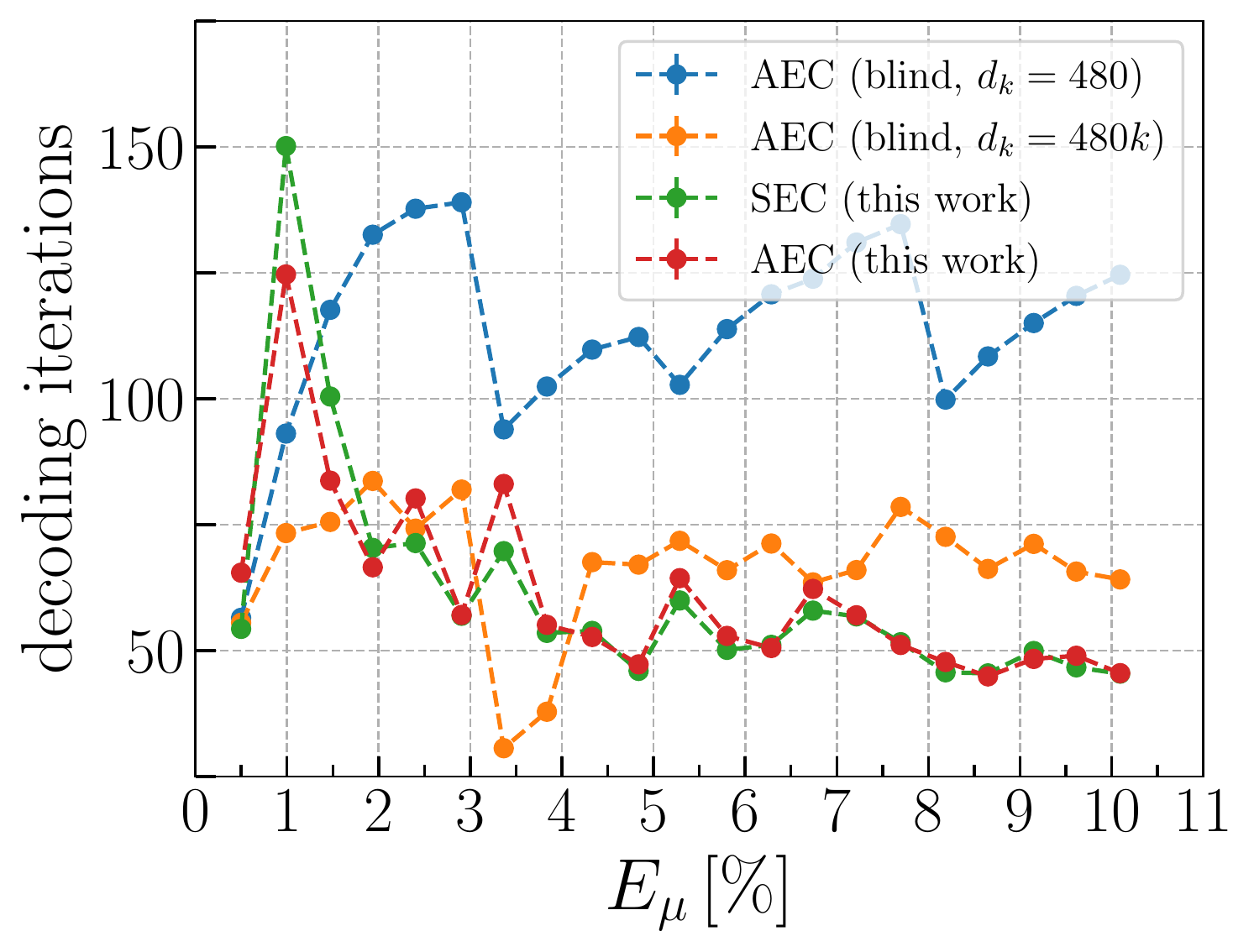}
    \caption{\footnotesize The dependencies of $f_\text{ec}$ and the average number of iterations, elapsed by the LDPC-decoder during one frame correction, on average signal pulse QBER $E_\mu$ for symmetric (SEC) and asymmetric error correction (AEC) approaches based on the simulated data analysis. The top axis represents the most frequently chosen LDPC code rate for the given QBER interval.}
    \label{fig:simulation}
    \label{fig:iters}
\end{figure*}

\begin{table}[t!]\centering
    \begin{tabular}{|c| c| c| c| c| c| c| c| c| c|}
        \hline
        $\mu$ & $\nu_1$ & $\nu_2$ & $p_\mu$ & $p_{\nu_{1,2}}$ & $\eta$ & $p_\text{dc}$ & $\tau_\text{dead}$ & $p_\text{opt}$ \\
        \hline
        $0.30$ & $0.09$ & $0.003$ & $0.50$ & $0.25$ & $0.13$ & $10^{-6}$ & $5\,\mu$s & $0.02$ \\
        \hline
    \end{tabular}
    \caption{\footnotesize The decoy-state BB84 setup parameters, used in both simulated and experimental data analysis: mean photon number per pulse of signal ($\mu$), week ($\nu_1$) and vacuum ($\nu_2$) decoy states, corresponding state generation probabilities $\{p_\mu,p_{\nu_1},p_{\nu_2}\}$, single-photon detector quantum efficiency $\eta$, dark count probability $p_\text{dc}$, dead time $\tau_\text{dead}$, and optical error probability $p_\text{opt}$.}
    \label{tab:params}
\end{table}

In order to analyze the proposed asymmetric error correction (AEC) algorithm and compare it with the improved symmetric (SEC) approach and other asymmetric schemes as well, we first generate numerous bit strings of raw key for various average signal QBER values, $E_\mu\in\{0.005,0.01,\dots,0.105\}$, using a theoretical model of the decoy-state BB84 protocol with parameters listed in Table~\ref{tab:params}.
The results of our simulations are presented in Figure~\ref{fig:simulation} where we plot the efficiency $f_\text{ec}$ \eqref{eq:f_ec} and the average number of decoding iterations as functions of the average QBER.
For reference, on the upper-left plot we also put the theoretical efficiency of a code with a fixed rate for a given $E_\mu$ interval, $f_\text{th}=(1-R)/h_2(E_\mu)$, which is the best efficiency that an EC scheme without rate-adaptive technique can achieve without frame correction failure.
Note that in our scheme $f_\text{ec}$ can not be smaller than the initial value $f_\text{start}$ due to additional EC rounds.
The important result of this work is that the proposed AEC scheme closely approaches the SEC efficiency for error rates $E_\mu\gtrsim2\%$. On the contrary, in the lower QBER region the AEC efficiency increases faster than SEC but still does not exceed by more than $5\%$. Also one can notice the increase of the $f_\text{ec}$ variance due to significant error fluctuations that induce the instability in the performance of any scheme with fast LDPC codes ($R\geq0.75$). Therefore, the fast codes have more strict requirements for the compliance of the selected set $\{R,p,s\}$ to an actual QBER of the frame, especially in an asymmetric scheme. In particular, it implies more precise \textit{a priori} QBER estimation for low $E_\mu$. In terms of the number of decoding iterations, AEC performs slightly better than SEC in the lower range $E_\mu<2\%$, while for $E_\mu\gtrsim2\%$ they demonstrate nearly the same results.

\begin{table}[t!]\centering
    {\small
    \begin{tabular}{|l|c|c|c|c|}
        \hline
        \multirow{2}{*}{EC scheme} & \textit{a priori} QBER & code rate & payload bits & \multirow{2}{*}{$d_k$} \\
        & estimation & adaptation & disclosure & \\
        \hline
        AEC (blind, fixed $d_k$) & \no & \yes & \no & 480 \\
        \hline
        AEC (blind, variable $d_k$) & \no & \yes & \no & $480k$ \\
        \hline
        AEC/SEC (this work) & \yes & \yes & \yes & Eq.~\eqref{eq:d_k_rule} \\
        \hline
    \end{tabular}
    \caption{\footnotesize Key features of various EC schemes.}
    \label{tab:EC_schemes_summary}
    }
\end{table}

\begin{table}[t!]\centering
    {\small
    \begin{tabular}{|c|c|c|c|c|c|c|}
        \hline
        & EC scheme & $\ell_\text{frame}$ & $\alpha$ & $R:[E_\mu^{\min},E_\mu^{\max}]$  & $N_\text{add}^{\max}$ & $\delta$ \\
        \hline
        \multirow{4}{*}{\begin{turn}{90}$d_k=\delta$\end{turn}} & \multirow{2}{*}{\cite{Martinez12blind,Martinez2013high}} & \multirow{2}{*}{2\,000} & \multirow{2}{*}{0.1} & $0.8:[0.01,0.035]$, $0.7:[0.02,0.06]$, & \multirow{2}{*}{$1-5$} & \multirow{2}{*}{$\frac{\alpha\ell_\text{frame}}{N_\text{add}^{\max}}$} \\
        & & & & $0.6:[0.04,0.09]$, $0.5 : [0.07,0.12]$ & & \\
        \cline{2-7}
        & \multirow{2}{*}{this work} & \multirow{2}{*}{32\,000} & \multirow{2}{*}{0.15} & $0.8:[0,0.03]$, $0.7:[0.03,0.05]$, & \multirow{2}{*}{$10$} & \multirow{2}{*}{480} \\
        & & & & $0.6:[0.05,0.08]$, $0.5:[0.08,0.11]$ & & \\
        \hline
        \multirow{4}{*}{\begin{turn}{90}$d_k=k\delta$\end{turn}} & \multirow{2}{*}{\cite{Liu20}} & \multirow{2}{*}{64\,800} & \multirow{2}{*}{0.1} & $0.8:[0.01,0.02]$, $0.6:[0.03,0.07]$, & \multirow{2}{*}{$4$} & \multirow{2}{*}{648} \\
        & & & & $0.5: [0.08,0.1]$ & & \\
        \cline{2-7}
        & \multirow{2}{*}{this work} & \multirow{2}{*}{32\,000} & \multirow{2}{*}{0.15} & $0.8:[0,0.03]$, $0.6:[0.03,0.08]$, & \multirow{2}{*}{$4$} & \multirow{2}{*}{480} \\
        & & & & $0.5:[0.08,0.11]$ & & \\
        \hline
    \end{tabular}
    }
    \caption{\footnotesize The list of parameters for AEC schemes with fixed ($d_k=\delta$) and increased ($d_k=k\delta$) number of disclosed bits during additional rounds. In this work we slightly modify some QBER intervals from Refs.~\cite{Martinez12blind,Liu20} to cover the entire QBER range, and use maximum total number of reconciliation rounds $N_\text{add}^{\max}$, which is expected to show better $f_\text{ec}$. For fixed $d_k$ doubled $N_\text{add}^{\max}$ is used. Code rates and $\ell_\text{frame}$ in Ref.~\cite{Martinez2013high} insignificantly differ from those in Ref.~\cite{Martinez12blind}.}
    \label{tab:original_scheme_pars}
\end{table}

The basic principle of our method is a union of adaptive code rate selection and strategy of additional rounds organization. It could be potentially applied to the code represented by arbitrary parity check matrix (PCM) with any $\ell_\text{frame}$ and $\alpha$ values. We compare our EC with two asymmetric blind IR schemes that use different data disclosure rules for additional reconciliation rounds using common codes pool with fixed $\ell_\text{frame}=32\,000$ bits in order to provide fair comparison. Key features of the compared approaches are summarised in Table~\ref{tab:EC_schemes_summary}. In the AEC scheme, proposed in \cite{Martinez12blind,Martinez2013high}, in each additional round the number of disclosed bits is equal and fixed, $d_k=\delta$. In another AEC scheme from Ref.~\cite{Liu20} it is increased with iterations, $d_k=d_{k-1}+\delta$. In both schemes no initial shortened bits are generated and only punctured bits are used in additional rounds. Thus, in these two schemes $s=0$ and $p=\alpha\ell_\text{frame}$. One has to emphasize that the results obtained with these two methods are highly dependent on initial settings like LDPC code frame length, the quality of parity check matrices, amount of punctured bits, etc. 
Therefore, we set our implementations of these methods maintaining the original number of additional rounds (or larger), which in fact defines maximum $f_\text{ec}$ and FER for any blind method. 
We set the maximum number of additional rounds equal to 10 and 4 for AEC with fixed and variable steps respectively. This in turn results to $\delta=480$ bits. The parameter list of the original works and our adaptation to our LDPC setup is presented in Table~\ref{tab:original_scheme_pars}. One can see from Figure~\ref{fig:simulation} that our AEC efficiency performs significantly better than both blind AEC schemes in the entire QBER region.


\begin{figure*}[t!]\centering
    \includegraphics[width=0.45\textwidth]{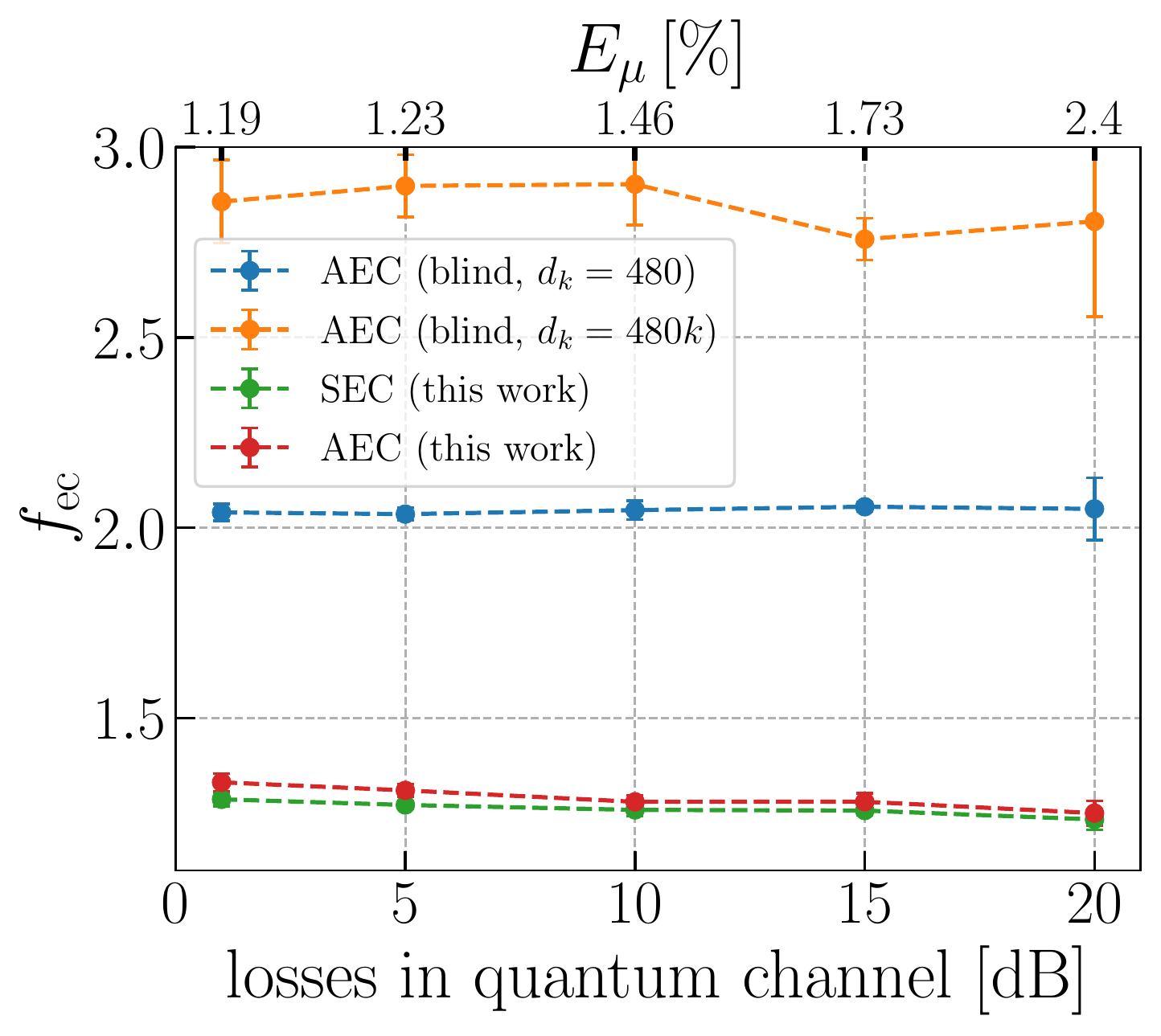}
    \includegraphics[width=0.45\textwidth]{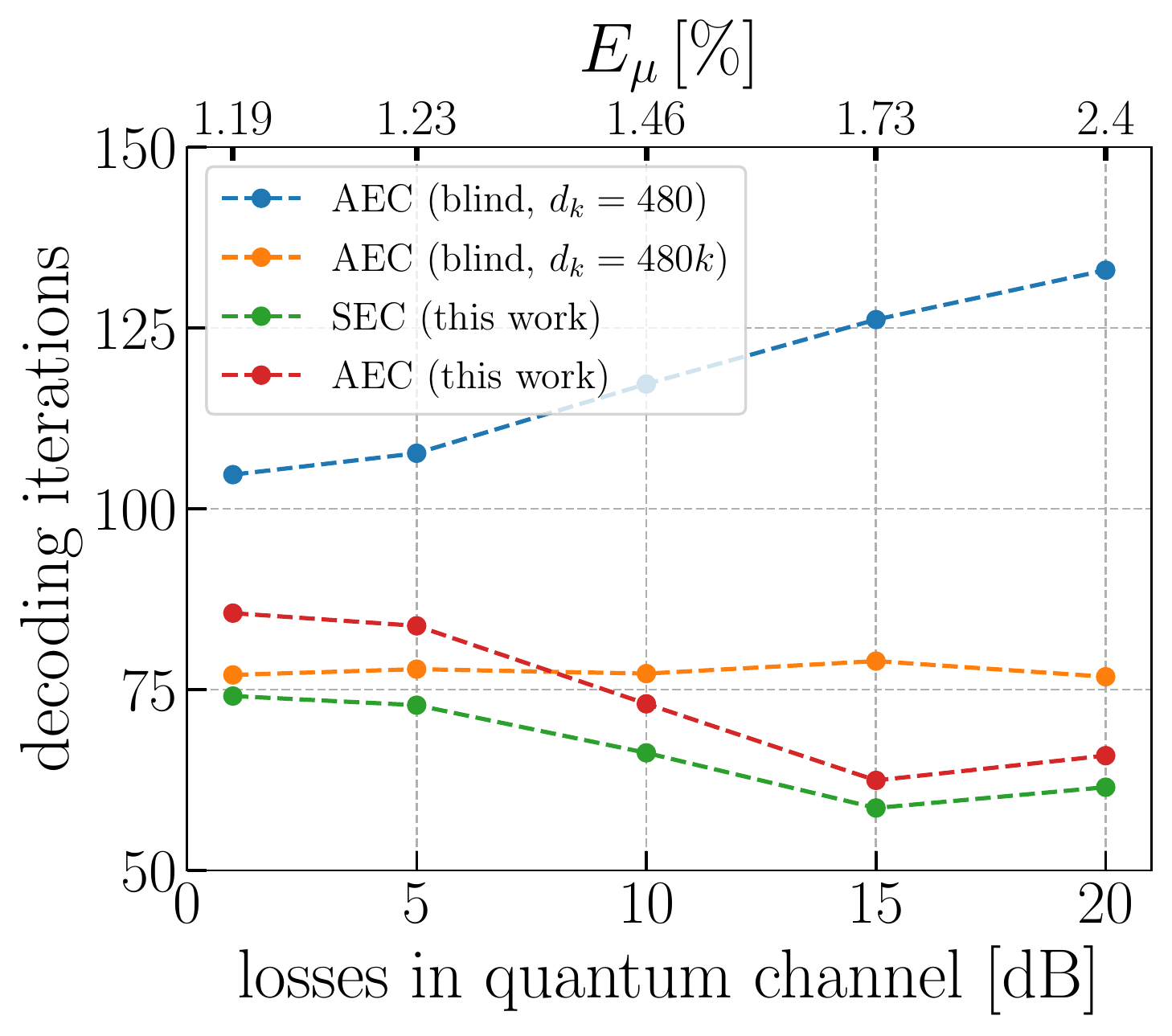}
    \includegraphics[width=0.45\textwidth]{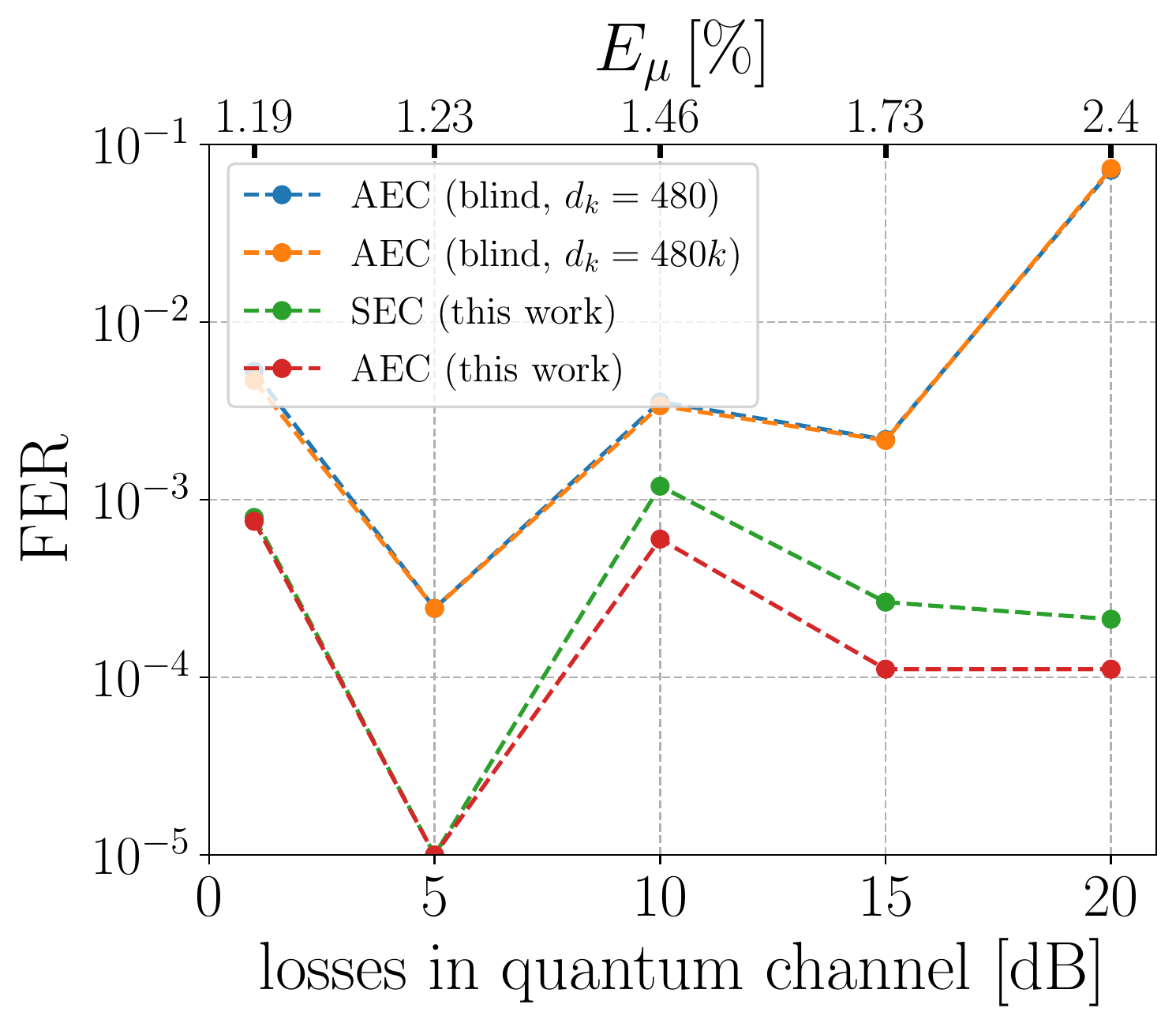}
    \includegraphics[width=0.45\textwidth]{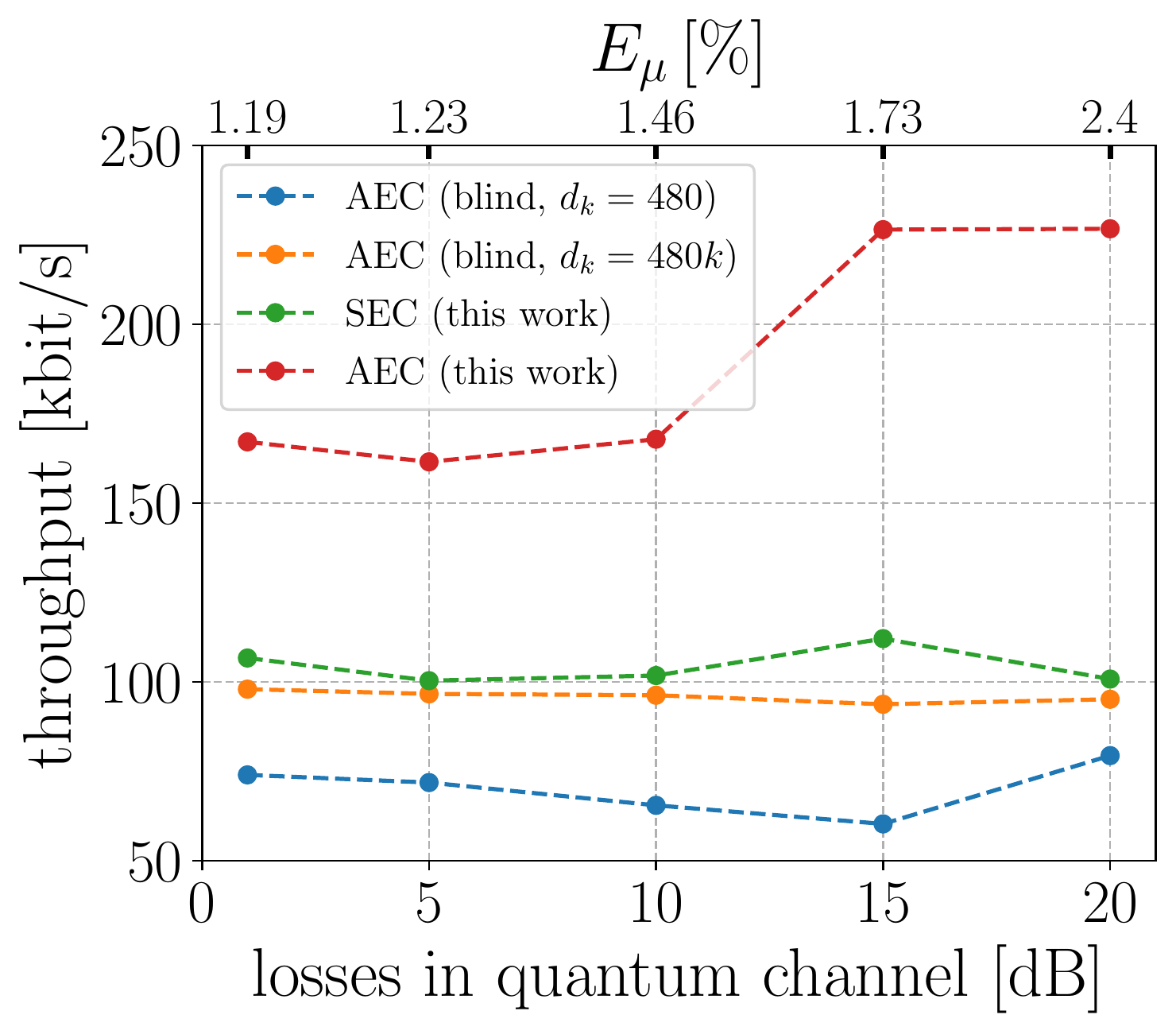}
    \includegraphics[width=0.45\textwidth]{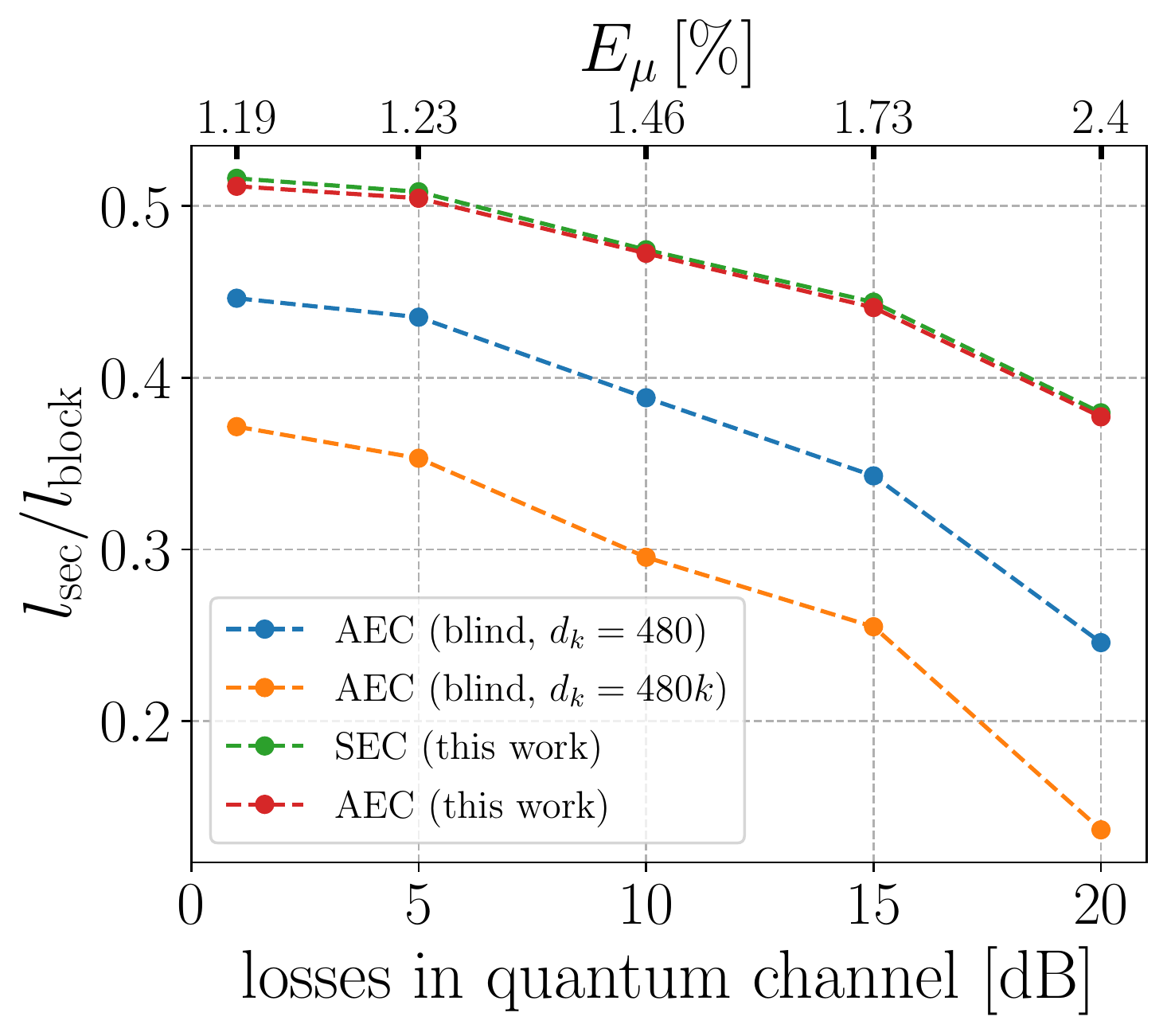}
    \includegraphics[width=0.45\textwidth]{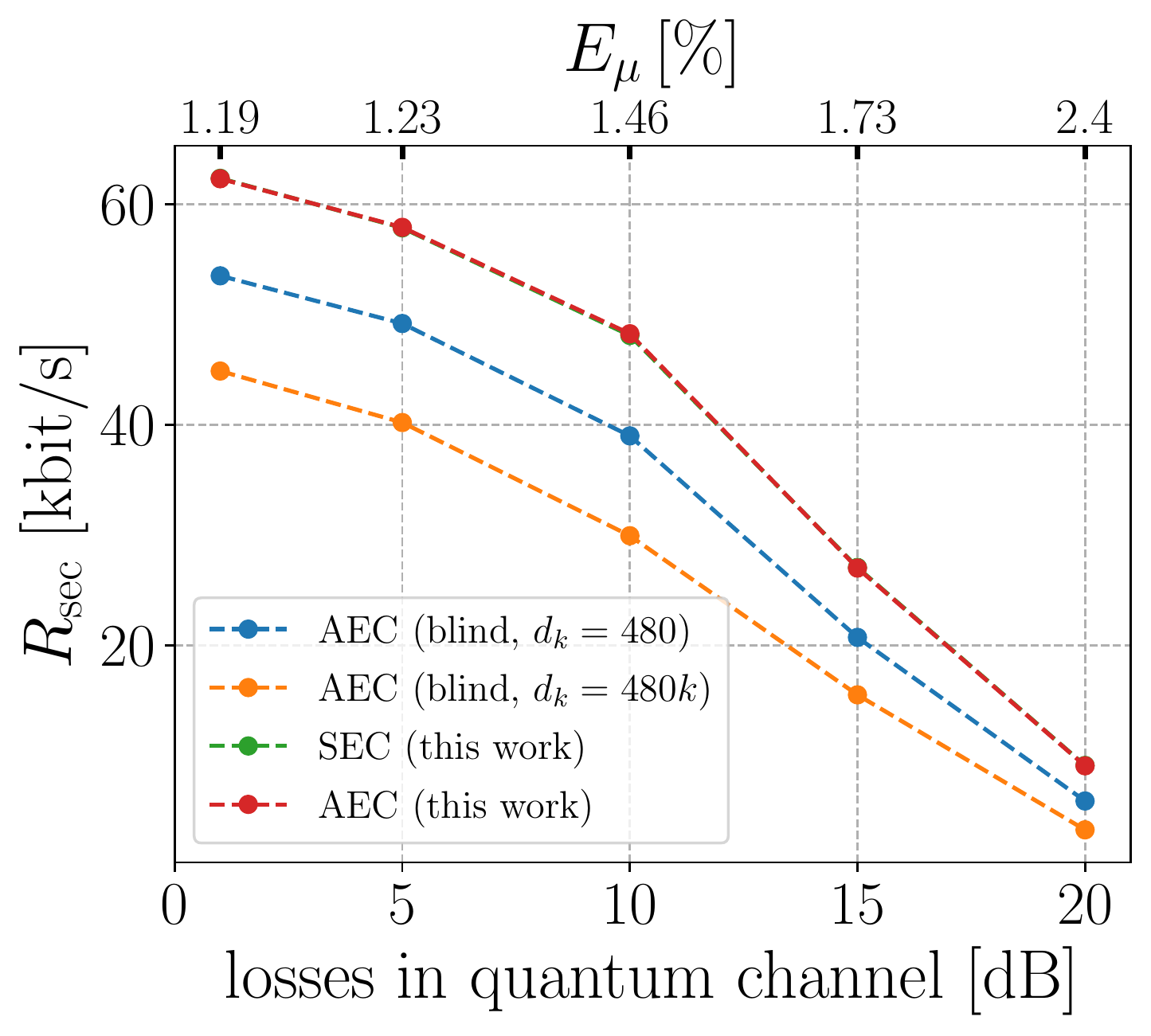}
    \caption{\footnotesize Performance of various error correction schemes on the real experimental data obtained with the industrial QKD devices by QRate.}
    \label{fig:dump}
\end{figure*}

Although the efficiency metric \eqref{eq:f_ec} is very informative when comparing different EC schemes, one has to consider another critical quantity of practical IR process -- the decoding time consumption that is proportional to the total number of decoder iterations. On the bottom plot in Figure~\ref{fig:iters} we show the average total number of iterations depending on the QBER level. This number evaluation includes both basic and additional reconciliation rounds. We have to mention that all results are obtained on a single-processor setup without parallel computing. One can see that the proposed adaptive AEC method is about two times faster than AEC with fixed step and slightly faster than AEC with variable step for $E_\mu\gtrsim4\%$.

Finally, we test the real EC performance on experimental data, obtained with industrial QKD devices manufactured by QRate~\cite{Kiktenko17_2} for various losses in the quantum channel up to 20\,dB (100\,km@0.2\,dB/km). Our results are presented in Figure~\ref{fig:dump}. The upper plots show similar behaviour as in Figure~\ref{fig:simulation}: the proposed AEC and SEC are very close in terms of efficiency, and SEC is slightly faster than AEC in terms of decoder iterations. The AEC scheme with fixed step is less efficient and slower in the entire loss range. Although AEC with variable step is a bit faster for losses up to 7\,dB, but it has much worse efficiency.

Having a lot of experimental data in our disposal (45,000 frames per dot), we also study the frame correction failure probability. One can see from Figure~\ref{fig:dump} that FER of our AEC scheme is less than $10^{-3}$ what is about one order of magnitude smaller than FER of AEC with fixed/increased step which can reach 10\% at 20\,dB. Also one can point out that the failure probability of the AEC scheme is slightly better than of SEC for losses starting from 5\,dB.

For practical QKD setup the IR throughput has to be analyzed as well \cite{Martinez2013high}. The throughput defines how many bits per second the EC algorithm can proceed and depends on two basic factors. The first one is the decoder's iteration cost, i.e. time spent on one belief propagation algorithm execution, determined by CPU performance, number of threads and chosen code rate. 
The second factor is the total number of iterations in all rounds, which increases with additional rounds.
Both blind AEC approaches have the lowest throughput because of their high number of iterations and FER. The SEC approach needs smaller number of additional rounds since the knowledge of the smallest LLR positions leads to faster convergence of the decoder. However, the cost of each decoder iteration overweights the number of rounds and leads to lower throughput in comparison to our AEC approach. For these reasons the developed AEC scheme demonstrates about two times better throughput with respect to the three other schemes.

On the bottom plots in Figure~\ref{fig:dump} we present the overall result of all previously discussed metrics and effects -- the normalized secret key length and generation rate. One observes that AEC and SEC have almost identical results and gain enhancement of about 20\%(40\%) in $\ell_\text{sec}$ and $R_\text{sec}$ with respect to AEC with fixed(variable) step. This fact clearly confirms the advantage of our AEC approach over the previously studied blind AEC versions. Another very important conclusion is that the introduced AEC algorithm demonstrates practically the same or sometimes even better performance than SEC.

\section{Conclusions}\label{sec:conclusion}

In this work we suggest a new approach to asymmetric error correction that could be used in practical QKD systems with limited computational resources on one side. We take the symmetric blind information reconciliation \cite{Kiktenko17sym} as a basis and propose such improvements as \textit{a priori} QBER estimation, different code rate selection and punctured bits disclosure rule. In particular, using the exponential moving average QBER of the previous verified frame together with decoy-state QBER allows the algorithm to detect gradual error rate changes and sudden bursts as swell and quickly adapt the code. Novel \textit{a priori} error estimation method we efficiently apply together with slightly changed rate-adaptive technique and blind-like interactive information reconciliation. Then, for the first time, we apply these features in the asymmetric approach. To compare various schemes we study several EC performance metrics and the secret key length/rate as final benchmarks on simulated and real data. We find that the improved symmetric and new asymmetric schemes demonstrate very close efficiency and the average number of decoding iterations in rather wide QBER range ($E_\mu\gtrsim2\%$). Thus a very important result of this work is that our asymmetric scheme turns out to be not inferior to the symmetric one neither in efficiency nor in secret key generation rate. We also make a comparison with two asymmetric blind schemes with the fixed and variable steps of the number of disclosed bits per additional round. We find that both proposed non-blind interactive approaches demonstrate a clear advantage over the blind ones. In this way one can conclude that the developed adaptive error correction approach can be efficiently used in decoy-state BB84 setups with fluctuating QBER level and asymmetric computational resource allocation.

\section{Acknowledgments}
This work is supported by the the Priority 2030 program at the National University of Science and Technology ``MISIS'' under the project K1-2022-027.

\normalem
\bibliographystyle{utphys}
\bibliography{bibliography_qkd.bib}

\end{document}